\documentclass[smallextended,natbib]{svjour3}
\smartqed
\usepackage{amsmath,amsfonts,amssymb,graphicx}
\def\gcd{\mathop{\mathrm{gcd}}}
\usepackage{longtable}
\usepackage{algorithm,algorithmicx,algpseudocode}

\def\proof{\noindent\textsf{Proof.}\ }

\usepackage{tikz}
\usetikzlibrary{automata,positioning}

 \journalname{ }
\begin{document}
\title{Universal evolutionary model for periodical species}

\author{Eric Goles \and  Ivan Slapni\v{c}ar \and  Marco A. Lardies}%
\authorrunning{Goles, Slapni\v{c}ar and Lardies}
 
\institute{E. Goles \at 
Facultad de Ingenieria y Ciencias, Universidad Adolfo Iba\~nez, Diagonal
Las Torres 2700, Pe\~nalolen, Santiago, Chile \email{eric.goles@uai.cl.}\and
I. Slapni\v{c}ar \at corresponding author, 
University of Split, Faculty of Electrical Engineering, Mechanical Engineering and Naval Architecture, 
R.\ Bo\v{s}kovi\'{c}a 32, 21000 Split, Croatia \email{ivan.slapnicar@fesb.hr} \and
M. Lardies \at Facultad de Artes Liberales, Universidad Adolfo
  Iba\~nez, Santiago, Chile \email{marco.lardies@uai.cl}
  } 
  
\date{Received: date / Accepted: date}

\maketitle

\begin{abstract}
Real-world examples of periods of periodical species range from cicadas whose life-cycles are larger prime numbers, like 13 or 17, to bamboos whose periods are large multiples of small primes, like 40 or even 120.
The periodicity is caused by interaction of species, be it a predator-prey relationship, symbiosis, commensialism,  or competition exclusion principle.  We propose a simple mathematical model which explains and models all those principles, 
including listed extremel cases. This, rather universal, {\em qualitative} model is based on the concept of a local fitness function, where a randomly chosen new period is selected if the value of the global fitness function of the species increases.
Arithmetically speaking, the different interactions are related to only four principles: given a couple of integer periods either (1) their greatest common divisor is one, (2) one of the periods is prime, (3) both periods are equal, or (4) one period is an integer multiple of the other.

\end{abstract}

\section*{Declarations}

\subsection*{Funding} 
Not applicable.

\subsection*{Conflicts of interest}
The authors declare that they have no conflict of interest.

\subsection*{Availability of data and material}
Not applicable. 

\subsection*{Code availability} Julia program used to run simulations and produce figures  is available at
\texttt{https://github.com/ivanslapnicar/EvolutionaryModel.jl}.

\subsection*{Authors' contributions}
E.\ G.\ conceived of project. I.\ S.\ computed simulations. 
E.\ G., I.\ S.\ and M.\ L.\ carried out research and wrote the manuscript.

\subsection*{Acknowledgments}
Part of this work was done while E.\ G.\ was visiting University of Split. Part of this work was done while 
I.~S.~was visiting Universidad Adolfo Iba\~nez supported by the grant from Centro de modelamiento matematico, Santiago, Chile. 
We would like to thank James Powell, Utah State University, Logan, for commenting on the final draft of the paper. 

\allowdisplaybreaks
\section{Introduction}\label{sec:introduction}

Life history diversity is a remarkable feature of living species and underlies fundamental evolutionary questions \cite{Rof02}. Periodical species are well-known for their mass and synchronous reproduction \cite{Jan76}. The use of an appropriate model to evaluate synchronicity in periodical species depends first on the life-history strategies of the species. For example, some species have a life-history with discrete non-overlapping generations, that is, there are no adult survivors from one generation to the next. Such species includes annual plants, annual insects, salmon, periodical cicadas and bamboos species \cite{Roc15}. Once the adults reproduce, they perish, and the future of the population is based on the dormant or juvenile stage of the organism. In periodical species, the population emerges synchronously from the ground or benthos (cicadas, bamboos, red or brown tides of phytoplankton species) as juveniles or adults. The growth rates before emerging in periodical cycle species are obviously affected by environment being the main driver for maintaining a minimum viable population (for example, extinction probability) the interspecific interactions with predators (including parasitoids), symbiosis, interspecific competition and/or disturbances in the habitat \cite{LlDy66,May83,Gra05,BPB18}.
Furthermore, the nature of these interactions can vary depending on the evolutionary context and environmental conditions in which they occur \cite{Ric08},
nevertheless the synchrony is maintained despite abrupt changes in the environment \cite{Che73,KoLi13}.

Understanding that natural forces drive this extraordinary periodical cycles is a central question in ecology. To understand these evolutionary processes, a broader view is needed of the properties of multiscale spatio-temporal patterns in species–environment interactions \cite{GRS08}.
Periodical species are plants or animals which emerge in nature every $T>1$ years, like some few species of cicadas (every 3, 13, 17 years) or bamboos (2, 3, 5, 15, 32, 60, 120 years), and other flowering plants \cite{KYM11}.
Density-dependent selection is the simplest form of feedback between an ecological effect of an organism’s own making (crowding due to sustained population growth) and the selective response to the resulting conditions \cite{TLH13}.
Density-dependent fitness and density-dependent selection are critical concepts underlying ideas about adaptation to biotic selection pressures and the coadaptation of interacting species. To understand those periods (and other related processes) several models have been proposed \cite{TYS09}.
In most of them one of the relevant assumptions is the “satiate hypothesis”, that is, organisms have to emerge in very large numbers in order to satiate predators (birds, rats, etc.) and consequently linked to climate conditions \cite{KeSo02,KYM11} but also \cite{KoLi13}.
Those assumptions explains synchronicity but not necessarily $T>1$ 
life-cycles. Since the usual life-cycle of those organisms is annual, the question is how it may evolve to a $T>1$ period? Several evolutionary pressures have been proposed to explain periodical life-cycles.
In \cite{Yos97,YHT09} authors present a deterministic discrete populational model as a mechanism which is related with systematic low temperatures in the ice ages in Pleistocene epoch. In this case, under the cold temperature pressure, selected cicadas are those with large periods underground during the metamorphosis stage in order to avoid low temperatures (but see also
\cite{IKU15}). 
Other theories that promote stabilising selection provide explanations for the existence of synchronous seeding in bamboos \cite{VND15}.
From this later assumption they proposed a model such that from the usual synchronously annual emergence of seeds, a mutation with period $T>1$ may appear. However, some problems remain:  why periodical cicadas have prime life-cycles?; why bamboos have some small prime cycles as well as very large non-primes life cycles? Some authors have proposed a very simple model for cicadas prime life-cycles by assuming the existence of a periodical predator \cite{GSM01}.
In this context, it was proved that the fixed points for the non-extinct periodical cicadas are necessarily prime numbers. Nevertheless, simulations were done which exhibit convergence to prime numbers only by accepting mutations in a narrow temporal scale (for instance plus or minus one year concerning the current life-cycle). Other models take into account an hybridization hypothesis, that is, if the life-cycle of two species of prey are prime numbers, $T$ and $T’$ so they encounter only every multiple of $T\times T’$ years, this  diminishes drastically the probability of hybridization which will produce other life-cycles with less offspring
\cite{Yos97}.
However, it is important to remark that in order to satisfy the hybridization hypothesis it is enough to consider only relative primes life-cycles
\cite{TYS09}.
Based on the several gaps of the models presented previously, it is required to develop a mathematical theory that includes the possible mechanism of how periodical life-cycles are shaped by diverse kinds of  interspecific interactions in an evolutionay context in nature. 

	In this work we propose a universal qualitative framework for species with synchronic periodical cycles, which is based on local fitness functions and the evolution of species interaction evaluating the extinction probability of species. Additionally, we describe the convergence of the proposed model and completely characterize its fixed points, leading to conclusion that different interactions are related to only four principles. 
The paper is organised as follows: in Section 2.1 we define general model based on the notion of a fitness function. In Section 2.1 we classify all fitness functions which appear in our model and reduce them to 18 cases (fitness tuples)  which need to be studied. In Section 2.3 we describe the dynamics of the model and state the evolution algorithm. In Section 2.4 characterize all fixed points of the studied fitness functions in the case of bounded mutations In Section3 we present numerical simulations for two cases of bounded mutations as well as more general ``quantitative'' fitness functions. In Section 4 we discuss our results and 
some open questions, incluiding small modifications which exhibit  fixed points which are very large periods. In Section 5 we draw our conclusions.

\section{Model}\label{sec:model}

\subsection{General model}

Consider two species, $\mathcal{C}_1$ and $\mathcal{C}_2$, with a emergence period
$c_1,c_2\in \mathbb{N}$, $c_1,c_2\geq 2$  (in years).
Consider the interval $T=c_1\cdot c_2$, and the emergence functions
$$\chi_i : [1,T]\to \{0,1\},$$ such that
$\chi_i(t)=1$ if and only if the species $\mathcal{C}_i$ emerges in year
$t$.
Clearly, species $\mathcal{C}_1$ emerges exactly $c_2$ times in the interval
$[1,T]$ and, respectively,  $\mathcal{C}_2$ emerges exactly $c_1$ times.

Let us now define the (local) fitness functions associated to
$\mathcal{C}_1$ and  $\mathcal{C}_2$ as:
$$
f_i:\{0,1\}\times \{0,1\}\to \{-1,0,1\}, \quad i=1,2
$$
such that $f_1(0,*)=f_2(*,0)=0$,
that is, if species is dormant, the value of its fitness function is zero.
If the species is emergent, the value of its fitness function with respect to the other species may
be $1$, $0$ or $-1$ (good, neutral or bad).
This simple model is qualitative, but it captures all intersting types of behaviour. 

For given periods $c_1$ and $c_2$, the global (cumulative) fitness function
over the interval $[1,T]$ is obtained by simply
adding yearly values of local fitness function and dividing them by the number of years in which the
species appears in this interval:
\begin{align*}
  F_1(c_1,c_2)&=\frac{1}{c_2} \sum_{t=1}^T f_1(\chi_1(t),\chi_2(t)),\\
  F_2(c_1,c_2)&=\frac{1}{c_1} \sum_{t=1}^T f_2(\chi_1(t),\chi_2(t)).
\end{align*}

\subsection{Classes of fitness functions}

Given two species, $\mathcal{C}_1$ and $\mathcal{C}_2$, and their local fitness
functions, $f_1$ and $f_2$, respectively, the only interesting situations are
when the species are emergent. Therefore, the only
cases which need to be considered are given by the $4$-tuple
\begin{align*}
v\equiv (\nu_1,\nu_2,\nu_3,\nu_4)=(f_1(1,0),f_2(0,1),f_1(1,1),f_2(1,1))\in \{-1,0,1\}^4.
\end{align*}
We shall call such $4$-tuple \textit{fitness tuple} or just \textit{tuple}.
Consequently, we have a total of $3^4=81$ possibly different
tuples $v$.  We reduce this number in three steps to only 18 tuples that need to be studied as follows.  
First, we eliminate equivalent tuples\footnote{Two tuples are equivalent is they are obtained by symply exchanging species $\mathcal{C}_1$ and $\mathcal{C}_2$, that is  
$(\nu_1,\nu_2,\nu_3,\nu_4)\equiv (\nu_2,\nu_1,\nu_4,\nu_3)$.}.
After eliminating equivalent tuples, we are left with  $v$ to 45 tuples, which we divide into four classes \footnote{For the class $D$  we also take into account the symmetry when one exchanges species, that is, $(a,b,c,d)=(b,a,d,c)$.}:
\begin{align*}
A&=\{(a,a,b,b)\mid a,b\in \{-1,0,1\} \}\\
&=\{(-1,-1,-1,-1),(-1,-1,0,0),(-1,-1,1,1),(0,0,-1,-1),\\
& \qquad  (0,0,0,0),(0,0,1,1), (1,1,-1,-1),(1,1,0,0),(1,1,1,1)\},
\\
B&=\{(a,a,-1,0),(a,a,-1,1),(a,a,0,1)\mid a \in \{-1,0,1\} \}\\
&=\{(-1,-1,-1,0),(-1,-1,-1,1),(-1,-1,0,1), (0,0,-1,0),\\
& \qquad (0,0,-1,1),(0,0,0,1), (1,1,-1,0),(1,1,-1,1),(1,1,0,1)\},
\\
C&=\{(-1,0,a,a),(-1,1,a,a),(0,1,a,a)\mid a \in \{-1,0,1\} \}\\
&=\{(-1,0,-1,-1),(-1,1,-1,-1),(0,1,-1,-1), (-1,0,0,0),\\
& \qquad (-1,1,0,0),(0,1,0,0),(-1,0,1,1),(-1,1,1,1),(0,1,1,1)\},
\\
D&=\{ (a,b,c,d) \mid (a\neq b)  \wedge (c \neq d)\}\\
&\qquad \{(-1,0,-1,0),(-1,0,0,-1),(-1,0,-1,1),(-1,0,1,-1),(-1,0,0,1),\\
&\qquad (-1,0,1,0),(-1,1,-1,0),(-1,1,0,-1),(-1,1,-1,1),(-1,1,1,-1),\\
& \qquad (-1,1,0,1),(-1,1,1,0),(0,1,-1,0), (0,1,0,-1),(0,1,-1,1),\\
& \qquad (0,1,1,-1),(0,1,0,1),(0,1,1,0)\}.
\end{align*}

In the second step, we eliminate tuples without biological interest or meaning:
\begin{align*}
(-1,-1,-1,-1), (-1,-1,0,0), (0,0,-1,-1),
(0,0,0,0), (1,1,0,0),\\(1,1,1,1), (-1,-1,-1,0),
(-1,-1,-1,1),(-1,-1,0,1),(0,0,-1,0), \\(0,0,0,1),
(-1,0,-1,-1), (-1,1,-1,-1),(-1,0,0,0), (-1,1,0,0),\\(0,1,0,0),(0,1,1,1),
(-1,0,-1,0),(-1,0,-1,1),(-1,1,-1,0),\\(-1,1,-1,1),(0,1,0,1).
\end{align*}
For example, for the tuple $(-1,0,-1,0)$ the 
species does not have any positive evolution pressure and the other species is completely indifferent.

In the third step, we eliminate tuples for which one of the fitness functions is always negative according to the followinmg lemma:
\begin{lemma}
 For the tuples
 \begin{align*}
 & (0,1,-1,-1),\ (0,0,-1,1),\  (0,1,-1,1),\  (0,1,-1,0),\\
 & (0,-1,-1,0),\ (0,1,-1,0), \textrm{ and } (-1,0,1,-1)
 \end{align*}
 one of the global fitness functions is always negative.
\end{lemma}
\proof See Appendix \ref{ProofL1}.

Therefore, the final set of tuples to study, $\mathcal{V}$, consists of the 18 tuples listed in Table \ref{tab:1}. We call such tuples {\em relevant fitness tuples} and we
number them for further reference.
\begin{table}[hbtp]
\centering
\begin{tabular}{lll}
$v_1=(-1,0,1,1)$ & 
$v_2=(-1,1,1,1)$ &
$v_3=(0,1,1,1)$ \\
$v_4=(0,0,1,1)$ &
$v_5=(-1,-1,1,1)$ &
$v_6=(1,1,-1,-1)$ \\
$v_7=(1,1,-1,0)$ &
$v_8=(1,1,1,-1)$ & 
$v_9=(-1,1,1,-1)$ \\
$v_{10}=(-1,1,1,0)$ &
$v_{11}=(1,0,-1,0)$ &
$v_{12}=(0,1,1,-1)$ \\
$v_{13}=(0,1,1,0)$ &
$v_{14}=(1,1,1,0)$ &
$v_{15}=(0,-1,1,0)$ \\
$v_{16}=(-1,0,1,0)$ &
$v_{17}=(1,-1,-1,0)$ &  
$v_{18}=(1,-1,1,0)$ 
\end{tabular}
\caption{Relevant fitness tuples.}
\label{tab:1}
\end{table}

\begin{figure}[hbtp]\label{fig:graph}
\begin{center}
\begin{tikzpicture}%
  [>=stealth,
   shorten >=1pt,
   node distance=3cm,
   on grid,
   auto,
   every state/.style={draw=black!40, fill=black!5, very thick},
   baseline=-1.0ex,
  ]
\node[state] (mid)                   {$\mathcal{C}_1$};
\node[state] (right)[right=of mid]   {$\mathcal{C}_2$};
\path[->]
   (mid)  edge[bend right=20]    node[below]                     {$f_2(1,1)$} (right)
	  edge[loop above]    node                      {$f_1(1,0)$} (right)
   (right)	edge[bend right=20]	node[above]		  {$f_1(1,1)$} (mid) 
		edge[loop above]    node                      {$f_2(0,1)$} (right)
   ;
\end{tikzpicture}
\end{center}
\caption{Generic ecological graph.}
\end{figure}
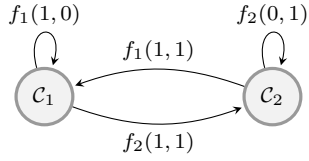

Each tuple may be represented by its respective ecological graph as in Figure \ref{fig:graph}. By inspecting corresponding ecological graphs, we classify tuples from Table \ref{tab:1} in four classes. The first class, $\mathcal{V}_1$, is the set off all tuples for which the corresponding graph is not strongly connected:
\begin{align*}
\mathcal{V}_1& =\{ v \in\mathcal{V} \mid (f_1(1,1)=0) \vee (f_2(1,1)=0) \}= \\ &=\{v_7,v_{10},v_{11},v_{13},v_{14},v_{15}, v_{16},v_{17},v_{18} \}.
\end{align*}
The other three classes are:
\begin{align*}
\mathcal{V}_2&=\{ v\in\mathcal{V} \mid f_1(1,1) = f_2(1,1)=1 \}
=\{v_1,v_2,v_3,v_4,v_5 \},\\
\mathcal{V}_3&=\{ v\in\mathcal{V} \mid f_1(1,1) = f_2(1,1)=-1 \} =\{v_6 \},\\
\mathcal{V}_4&=\{ v\in\mathcal{V} \mid f_1(1,1)\cdot f_2(1,1)=-1 \} =\{v_8,v_9,v_{12} \}.
\end{align*}

\subsection{Dynamics of the model}

We describe the  evolution of species $\mathcal{C}_1$ and $\mathcal{C}_2$.
Given initial periods $c_1,c_2\in\mathbb{N},\ c_1,c_2\geq 2$, the first iteration of the evolution game is as follows:
for the species $\mathcal{C}_1$ we propose random local change of its period $c_1$ to 
new period $c_1'\in\mathbb{N},\ c_1'\geq 2$. If the value of the global fitness function of the species $\mathcal{C}_1$ has improved, that is,  if 
$F_1(c_1',c_2)>F_1(c_1,c_2)$, the period $c_1'$ is accepted as the new period, $c_1\gets c_1'$, and the value of the global fitness function of the other species, $\mathcal{C}_2$, is recomputed.
The analogous procedure is now applied to the species $\mathcal{C}_2$.
Such iterations are repeated until there is no change. The pseudocode of the algorithm is given in Algorithm \ref{alg:evolution}.

\begin{algorithm}[hbtp]
    \caption{Evolution dynamics}
    \label{alg:evolution}
    \begin{algorithmic} 
        \Procedure{Evolution}{$c_1,c_2,v$} \Comment{$c_1,c_2\in\mathbb{N},\ c_1,c_2\geq 2$}
        \State $T=c_1\cdot c_2$
        \State compute $F_1(c_1,c_2)$ and $F_2(c_1,c_2)$
            \Loop
		\State choose $c_1'$ at random \Comment{$c_1'\geq 2$}
                \State $T'=c_1'\cdot c_2$
                \State compute $F_1'(c_1',c_2)=\displaystyle\frac{1}{c_2}\sum_{t=1}^{T'} f_1(\chi_1(t),\chi_2(t))$
                \If{$F_1'(c_1',c_2)>F_1(c_1,c_2)$} \Comment {the fitness of $\mathcal{C}_1$ improved}
                \State $c_1\gets c_1'$ \Comment{accept the new period}
                \State recompute $F_2(c_1,c_2)$
                \EndIf
                \State choose $c_2'$ at random \Comment{$c_2'\geq 2$}
                \State $T'=c_1\cdot c_2'$
                \State compute $F_2'(c_1,c_2')=\displaystyle\frac{1}{c_1}\sum_{t=1}^{T'} f_2(\chi_1(t),\chi_2(t))$
                \If{$F_2'(c_1,c_2')>F_2(c_1,c_2)$} \Comment {the fitness of $\mathcal{C}_2$ improved}
                \State $c_2\gets c_2'$ \Comment{accept the new period}
                \State recompute $F_1(c_1,c_2)$
                \EndIf
            \EndLoop\label{evolutionloop}
        \EndProcedure
    \end{algorithmic}
\end{algorithm}

We will also assume that the evolutionary game changes each species in a mutation interval $[-p,p]=[-p,-p+1,\ldots,-1,0,1,\ldots,p]$, that is, the 
arbitrarily far moves do not exist in the algorithm. More precisely, the new periods $c_1'$ and $c_2'$ are chosen as
$$
c_1'=c_1+k,\quad c_2'=c_2+l,\quad k,l\in[-p,p].
$$
We say that a couple of periods $(c_1,c_2)$ for the given tuple $v$ is a $p$-\textit{fixed point} if and only if
 $\forall k,l\in[-p,p]$ 
$$
F_1(c_1+k,c_2)\leq F_1(c_1,c_2)
$$
and 
$$
F_2(c_1,c_2+l)\leq F_2(c_1,c_2).
$$
In this context, fixed points are related to the interval of possible time jumps. 
We also speak of \textit{global fixed point} when the above is true for any $p\in\mathbb{N}$. 

\subsection{Characterization of fixed points}

For the next results we will consider the evolution algorithm for $p=1$, that is, in $[-1,1]$ mutation space. In particular, by ``local fixed point'' we mean $1$-fixed point. The results for $p$-fixed points are simillar (see Numerical simulations and Discussion). 

The first results are related with the relations between the sets of fixed points in any of the classes $\mathcal{V}_i$, $i=1,2,3,4$.
For that, let us define the set of fixed points as:
\begin{align*}
\mathcal{A}_k&=\{(c_1,c_2) \mid c_1,c_2\in \mathbb{N}, \textrm{  and it is an fixed point }\\
&\qquad \textrm{ for the tuple } v_k\}.
\end{align*}
The relations between the sets are as follows:

\begin{lemma}\label{proposition0}
For the tuples in the class $\mathcal{V}_1$ we have:
  \begin{align*}
  & \mathcal{A}_7\subsetneq \mathcal{A}_{11},\quad  \mathcal{A}_{10}=\mathcal{A}_{13}\subsetneq \mathcal{A}_{14},\quad 
  \mathcal{A}_{15} \subsetneq \mathcal{A}_{16},\\
  &\mathcal{A}_{10} \subsetneq \mathcal{A}_{16},\quad \mathcal{A}_{17} \subsetneq \mathcal{A}_{11},\quad 
  \mathcal{A}_{17} \subsetneq \mathcal{A}_{18}.
  \end{align*}
For the tuples in the class $\mathcal{V}_2$ we have:
  $$
  \mathcal{A}_1= \mathcal{A}_{4}=\mathcal{A}_{5}\subsetneq \mathcal{A}_{2}=\mathcal{A}_{3}.
  $$
For the tuples in the class $\mathcal{V}_4$ we have
  $$
  \mathcal{A}_9= \mathcal{A}_{12}\subsetneq \mathcal{A}_{8}.
  $$
\end{lemma}

\proof See Appendix \ref{ProofL2}.

Using Lemma \ref{proposition0}, we eliminate tuples $v_3$, $v_4$, $v_5$, $v_{12}$ and $v_{13}$ from Table \ref{tab:1}.
We shall call the remaining tuples inside each class \textit{representative tuples}. 
Now we characterize the fixed points inside each class.

\begin{proposition}\label{proposition1}
 For the representative tuples in the class $\mathcal{V}_1$ the structure of fixed points is as follows:
 \begin{itemize}
  \item[--] For  $v_7$ there exist global fixed points such that $\gcd (c_1,c_2)=1$, and there exist infinitely many local ones.
  \item[--] For $v_{10}$ there exist infinitely many local fixed points, but no global ones. 
  \item[--] For $v_{11}$ there exist global fixed points such that $\gcd (c_1,c_2)=1$, as well as infinitely many local ones.
  \item[--] For $v_{14}$ there exist global fixed points such that $\gcd (c_1,c_2)=1$, as well as infinitely many local ones.
  \item[--] For $v_{15}$, the couples $(c_1,c_2)$ such that $c_1=c_2$ are global fixed points. Further, there exist infinitely many local ones.
  \item[--] For $v_{16}$, the  couples $(c_1,c_2)$ such that $c_2\mid c_1$ are global fixed points, and there exist infinitely many local ones.
  \item[--] For $v_{17}$ there are no global fixed points, but there are infinitely many local ones. For instance, couples $(c_1,c_2)$ such that $c_1$ 
  is a prime number and $c_2<c_1-1$ are local fixed points.
  \item[--] For $v_{18}$, the  couples $(c_1,c_2)$ such that $c_1\mid c_2$ are global fixed points, and there exist infinitely many local ones.
 \end{itemize}
\end{proposition}

\proof See Appendix \ref{ProofP1}.

\begin{proposition}\label{proposition2}
 For the representative tuples in the class $\mathcal{V}_2$ we have:
 \begin{itemize}
  \item[--] $v_1$ admits the global fixed points $(c_1,c_2)$ such that $c_1=c_2$. Further, there exist also infinitely many local fixed points.
  \item[--] For $v_{2}$,  any couple $(c_1,c_2)$ such that $c_2\mid c_1$ is a global fixed point. Further, there exist infinitely many local fixed points.
 \end{itemize}
\end{proposition}

\proof See Appendix \ref{ProofP2}.

\begin{proposition}\label{proposition3}
 For the class $\mathcal{V}_3=\{v_6\}$, every $(c_1,c_2)$ such that $\gcd(c_1,c_2)=1$, is a global fixed point. 
 There exist infinitely many local fixed points.
\end{proposition}

\proof See Appendix \ref{ProofP3}.

\begin{proposition}\label{proposition4}
 For the representative tuples in the class $\mathcal{V}_4$ we have:
 \begin{itemize}
  \item[--] $v_8$ admits global fixed points $(c_1,c_2)$ such that $\gcd(c_1,c_2)=1$. Further, it admits infinitely many 
  local fixed points.
  \item[--] $v_{9}$ only admits local fixed points: if $(c_1,c_2)$ is such that $c_2$ is prime and $c_2\notin \{c_1-1,c_1+1\}$, then it is a 
  local fixed point. Further, the species $\mathcal{C}_1$ disappears. Also, there exist other infinite families of local fixed points.  
 \end{itemize}
\end{proposition}

\proof See Appendix \ref{ProofP4}.

\section{Numerical simulations}\label{sec:numerical}

We ran simulations to illustrate the behavior of the representative $4$-tuples. 
A couple of periods $(c_1,c_2)$ is a $p$-fixed point 
under simulation if, after thirty steps, neither of the species changes its period. 
We plot fixed points using the following convention: If $(c_1,c_2)$ is \textit{not} an fixed point, the corresponding square on the plot is white, 
otherwise it is red if both species survive, or gray if the species $\mathcal{C}_1$ disappears. The link to the complete code used to run simulations is given in Declarations.

\begin{figure}[hbtp]
\includegraphics[width=0.85\textwidth]{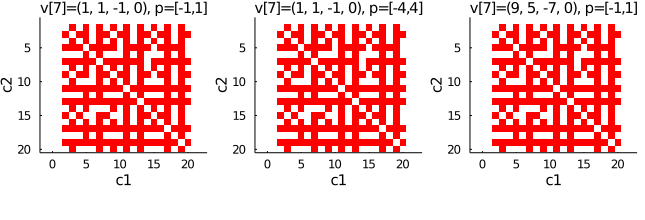}\\  
\includegraphics[width=0.85\textwidth]{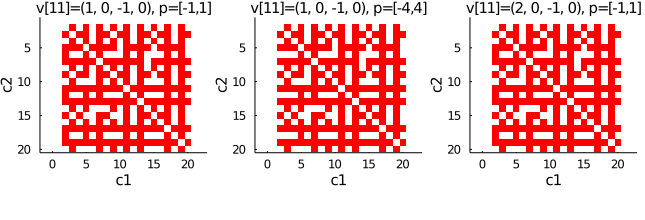} \\ 
\includegraphics[width=0.85\textwidth]{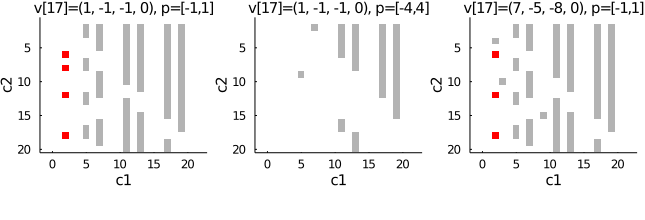}  
\caption{Attractors for tuples $v_7$, $v_{11}$ and $v_{17}$ from Proposition 1. The tuples  are of a “competition” type. Global fixed points for $v_7$ and $v_{11}$ are those with 
$\gcd(c_1,c_2)=1$. In 
$v_{17}$, the species $\mathcal{C}_1$ has a behaviour like periodic cicadas with prime number  fixed points. In this case, the species $\mathcal{C}_2$ disappears, since the fitness 
$F_2(c_1, c_2)$ is always negative. Attractors are marked with red squares if both species survive and with gray squares if one of the species disappears.}
\label{fig2}
\end{figure}

\begin{figure}[hbtp]
\includegraphics[width=0.85\textwidth]{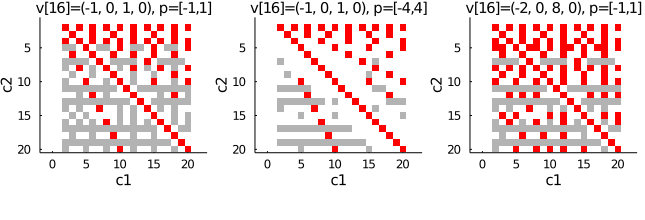}\\ 
\includegraphics[width=0.85\textwidth]{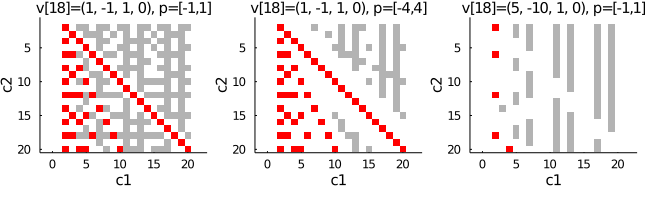}   
\caption{Attractors for tuples $v_{16}$ and $v_{18}$ from Proposition 1.
Here one of the global fitness functions is 
constant ($F_2(c_1, c_2) = 0$ for $v_{16}$ and $F_1(c_1, c_2) = 1$ for $v_{18}$), while the other  species has a bamboo-like behaviour -- it emerges as an integer multiple of the 
constant one. In both cases there is a “commensalism” interaction. }
\label{fig3}
\end{figure}

\begin{figure}[hbtp]
\includegraphics[width=0.85\textwidth]{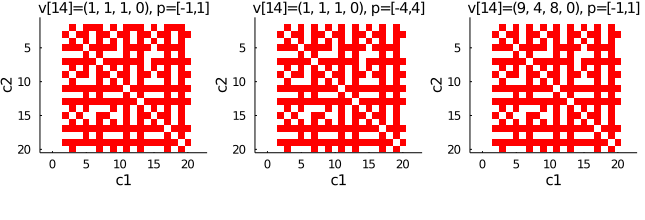} \\ 
\includegraphics[width=0.85\textwidth]{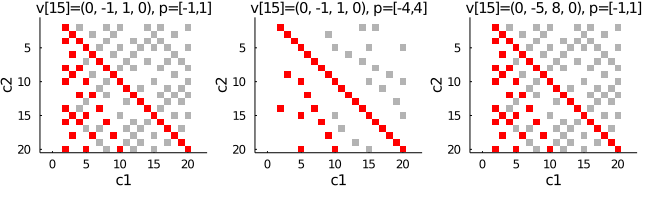} 
\caption{Attractors for tuples $v_{14}$ and $v_{15}$ from Proposition 1. For $v_{1}$, global fixed points are again those with $\gcd(c_1,c_2)=1$, and for  $v_{15}$, global fixed points are those with $c_1=c_2$.}
\label{fig4}
\end{figure}

\begin{figure}[hbtp]
\includegraphics[width=0.85\textwidth]{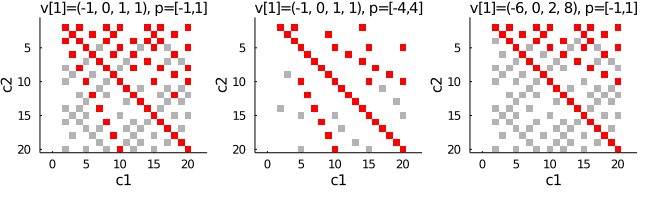}\\ 
\includegraphics[width=0.85\textwidth]{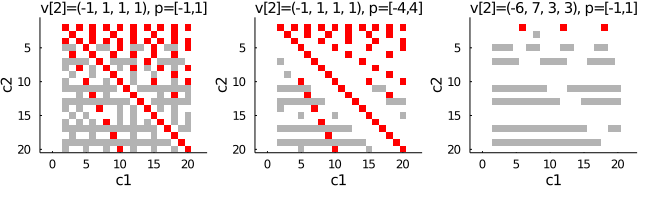} 
\caption{Attractors for tuples $v_1$ and $v_2$ from Proposition 2. 
The tuple $v_1$ generates symbiosis of the two species, that is, they 
have to appear at the same time (here global attractrs are again those with $c_1=c_2$).  The tuple $v_2$ generates a “commensalism” 
relation: the period $c_1$ has to be an integer multiple of $c_2$, $c_2\mid c_1$, 
independently of $c_1$ (a “bamboo” effect). }
\label{fig5}
\end{figure}

\begin{figure}[hbtp]
\includegraphics[width=0.85\textwidth]{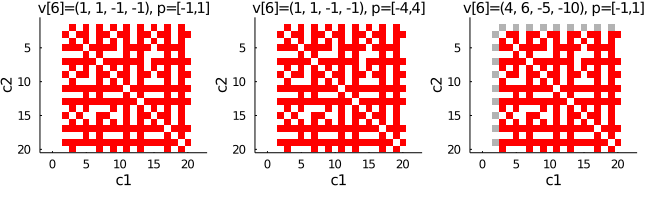} 
\caption{Attractors for tuple $v_6$ from Proposition 3.  This is the 
case of two species which cannot emerge together -- they are mutually harmful. In 
this sense, the competition exclusion principle can originate 
local extinction in populations. Global fixed points are those with $\gcd(c_1,c_2)=1$.
Notice that the fixed points for fitness tuples $v_6$, $v_7$, $v_8$, $v_{11}$ and $v_{14}$ are similar. }
\label{fig6}
\end{figure}

\begin{figure}[hbtp]
\includegraphics[width=0.85\textwidth]{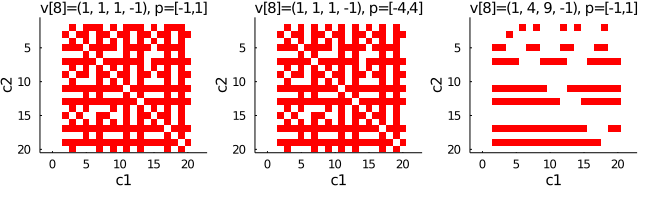} \\ 
\includegraphics[width=0.85\textwidth]{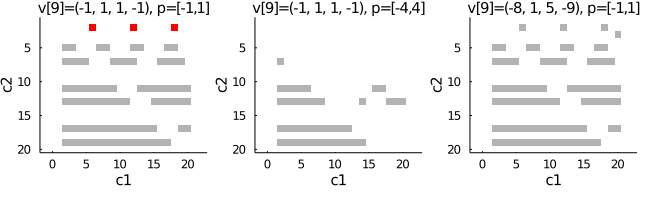} 
\caption{Attractors for tuples $v_8$ and $v_9$ from Proposition 4. Both tuples generate 
predator-prey interaction, the species $\mathcal {C}_2$ being the prey. 
For $v_8$, $\mathcal{C}_1$ does not depend on $\mathcal{C}_2$ (it may feed on other items), but $\mathcal{C}_2$ has to escape, hence global fixed points are those with $\gcd (c_1, c_2) = 1$. 
For $v_9$ there is dependence between the prey and the predator: 
the species $\mathcal{C}_1$ must eat $\mathcal{C}_2$ in order to survive, and 
$\mathcal{C}_2$ must escape from $\mathcal{C}_1$ in 
order to survive. Local fixed points are those with prime $c_2$ and $c_2\notin\{c_1-1,c_1+1\}$ (cicadas-like behaviour). Here also $F_1(c_1, c_2) < 0$, so the species $\mathcal{C}_1$ disappears. }
\label{fig7}
\end{figure}

\begin{figure}[hbtp]
\centering
\includegraphics[width=0.85\textwidth]{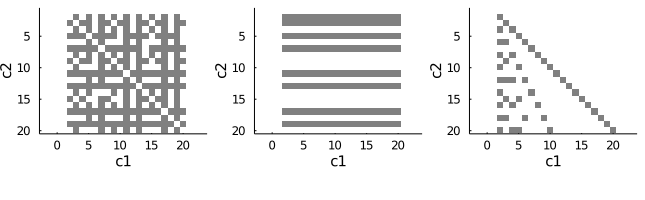}
\caption{For easier visual reference, here are canonical plots of three typical situations: gray denotes pairs $(c_1,c_2)$, $c_1,c_2 >1$, such that $\gcd(c_1,c_2)=1$,
$c_2$ is prime, or $c_2\mid c_1$, respectively.}
\label{fig8}
\end{figure}

The plots in Figures \ref{fig2}-\ref{fig7} depict characteristic behaviour of the qualitative model described in Propositions 1-4 (see also Discussion). For easier visual reference when interpreting the figures, Figure \ref{fig8} shows canonical plots of three typical situations.

The first plot in each row shows the results of the simulation for the corresponding fitness tuple for the mutation interval $[-p,p]=[-1,1]$. The second plot in each row shows the results for the mutation interval $[-p,p]=[-4,4]$ -- as expected, by increasing mutation interval some local fixed points cease to be fixed points, but global fitness points remain.  

We also simulated what happens in more ``qualitative'' setting, that is, when
local fitness functions are generalized such that
$$
f_i:\{0,1\}\times \{0,1\}\to \mathbb{R}, \quad \quad i=1,2.
$$
In order to run these tests, each entry of fitness tuple was multiplied by a random integer from the interval $[1,10]$ and the mutation interval was $[-p,p]=[-1,1]$.
The results are displayed in the third plot in each row. It is worth noting that the plots are indeed similar to the plots in the first two columns. These simulations indicates that our qualitative model covers well different qualitative cases, but this should by no means be considered a proof of such claim, without further analysis. Clearly, analysis for any particular fitness tuple could be done, if the specific case of interest appears.

\clearpage 

\section{Discussion}

By analyzing the ecological factors that shape development, reproduction and 
survival, life history theory seeks to explain the evolution of the major 
features of life cycles \cite{Ste92}. The complex spatial and temporal
populational dynamics of periodical species are largely consequences of 
evolution of species interactions. Our models are relatively simple and describe 
accurately the evolution of several types of periodical species with 
synchronized emergence including species associated with prime-numbered 
(i.e.\ insects) and non-prime numbered life-cycles (i.e.\ plants) \cite{HVS94,HaYa13,Gue14}.
Several fitness tuples in our model
show that various life cycles and their respective periodical emergences 
extinguish each other. On the other hand, our computer simulations demonstrate 
that the evolution of the periodicity is possible under restricted conditions. Here 
we first comment four propositions and their respective classes of fitness tuples.

 In the 
class $\mathcal{V}_1$ (Proposition \ref{proposition1}) there are two types of behaviour. The 
tuples $v_7$, $v_{11}$ and $v_{17}$ are of a “competition” type. In 
$v_{17}$, the species $\mathcal{C}_1$ has a behaviour like periodic cicadas with prime number 
fixed points (and in this case the species $\mathcal{C}_2$ disappears, since the fitness 
$F_2(c_1, c_2)$ is always negative). The other type of behaviour is the one 
generated by the tuples $v_{16}$ and $v_{18}$. Here one of the global fitness functions is 
constant ($F_2(c_1, c_2) = 0$ for $v_{16}$ and $F_1(c_1, c_2) = 1$ for $v_{18}$), while the other 
species has a bamboo-like behaviour -- it emerges as an integer multiple of the 
constant one. In both cases there is a “commensalism” interaction, a special 
kind of symbiosis. 

The tuples $v_1$ and $v_2$ from the class $\mathcal{V}_2$ (Proposition \ref{proposition2}) generate two types of behaviour. The tuple $v_1$ generates symbiosis of the two species, that is, they have to appear at the same time. The tuple $v_2$ generates a “commensalism” 
relation: the period $c_1$ has to be an integer multiple of $c_2$ (which is 
independent of $c_1$), so it is also a kind of a “bamboo” effect. 

The tuple $v_6$ (Proposition \ref{proposition3}) is interesting since it is the case of two species which cannot emerge together -- they are mutually harmful. In 
this sense, it is widely known that the competition exclusion principle can originate 
local extinction in populations \cite{Ben89} and could be represented by 
this proposition. According to the host-predator hypothesis, the parasitoids, 
predators and certain microorganisms specializing on immature stages of a 
species lack prey or a substrate in the year when only adults are present. The 
following year the predators are at low levels when the immature host numbers 
are again high. The same would hold for specialist predators of adult stage. 
Several mathematical models have been explored in an attemp to understand the 
origin of periodicity. Hoppenstadt and Keller \cite{HoKe76} showed that synchronized 
emergence of a single year class is a possible consequence of predation given a 
limited environmental carrying capacity and life cycle lengths above a certain 
threshold value. Bulmer \cite{Bul77} also demonstrated that predation alone will not 
cause periodical behavior except as the accidental result of particular initial 
conditions. In this sense, our model is concordant with the last reported case.

The tuples $v_8$ and $v_9$ from the class $\mathcal{V}_4$ (Proposition \ref{proposition4}) both generate 
predator-prey interaction, the species $\mathcal {C}_2$ being the prey. 
For $v_8$, $\mathcal{C}_1$ does not depend on $\mathcal{C}_2$ (it may feed on other items), but $\mathcal{C}_2$ has to escape, hence $\gcd (c_1, c_2) = 1$. 
For $v_9$ there appears real dependence between the prey and the predator: 
the species $\mathcal{C}_1$ must eat $\mathcal{C}_2$ in order to survive, and 
$\mathcal{C}_2$ must escape from $\mathcal{C}_1$ in 
order to survive. In this case, classes of fixed points are those with prime $c_2$ in 
which case $F_1(c_1, c_2) < 0$ so the species $\mathcal{C}_1$ disappears. 
This was the case studied widely in \cite{GSM00,GSM01}. 

\begin{figure}[hbtp]
\centering
\includegraphics[width=0.6\textwidth]{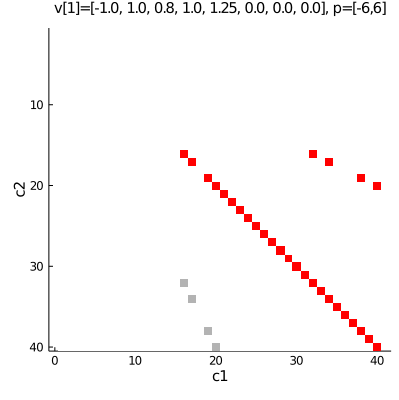}
\caption{Simulation using full fitness tuple with values outside of the set $\{-1,0,1\}$, generates periods longer than 12 for both species, and the species $\mathcal{C}_1$ has periods 32, 34, 38, 39 and 40, as observed in the population 
of long-period masting bamboos.}
\label{fig9}
\end{figure}

Some open questions and generalizations are related to considering full fitness 
tuples 
\begin{equation*}
v = (f_1(1, 0), f_2(0, 1), f_1(1, 1), f_2(1, 1), f_1(0, 1), f_2(1, 0), f_1(0, 0), f_2(0, 0)),
\end{equation*}
and allowing values outside of the set $\{-1, 0, 1\}$. For example, large dormant 
intervals of bamboos could be explained by the need to minimize energy (see \cite{GaBo70}), so in this case the “dormant” values $f_1 (0, 1)$ or $f_2 (1, 0)$ might 
be positive. In particular, our simulation for the fitness tuple $v = (-1, 1, 
0.8, 1, 1.25, 0, 0, 0)$ generates fixed points with periods $26$, $32$, $34$, $38$, $39$ and $40$ for the 
species $\mathcal{C}_1$ (see Figure \ref{fig9}), which were observed in the population of long-period masting bamboos, see \cite[Table 1]{Jan76} and 
\cite[Appendix, Table S1]{VND15}. Also, 
bamboos periodical life-cycles can be dramatically affected during the 
flowering event by anthropogenic habitat modification that cause delayed 
reproduction, see \cite{WFL16}. That is, the combination of a fixed 
juvenile development time and a long adult life could favor the development of 
periodical behavior but does not guarantee it, see \cite{LaHa86}. Again, this ``qualitative'' simulation is by no means proof or explanation of long period masting, it is an example of the universality of our theoretical model. 

\section{Conclusion}

To conclude, the presented theoretical model covers all types of periodic behaviour.
We learned that, from a simple arithmetic point of view, the 
different interactions are related to only four principles. Given a 
couple of integer periods $(c_1, c_2)$ the principles are:
\begin{itemize}
\item[--]
$\gcd(c_1, c_2) = 1$ (predation or interspecific competition interaction as for 
tuples $v_6$, $v_7$, $v_8$, $v_{11}$ and $v_{14}$),  
\item[--] one of $c_1$ or $c_2$ has to be a prime (cicadas-like behaviour as 
for $v_9$, $v_{10}$ and $v_{17}$),  
\item[--] $c_1 = c_2$ (symbiosis interaction as for $v_1$ and $v_{15}$), and 
\item[--] one of $c_1$ or $c_2$ has to be an integer multiple of the other (bamboos-like 
interaction as for $v_2$, $v_{16}$ and $v_{18}$). 
\end{itemize}
Some of these priciples are indeed observed in nature - periodic cycadas and long-period masting bamboos.

To be periodical, a species must have a fixed life cycle length and adults must 
appear synchronously, reproduce only once, and die \cite{HVS94}. We 
demonstrate that the emergence of synchrony in some species can occur by chance 
events which disrupt this bet-hedging strategy and set the stage for periodicity. 
Our mathematical model predicts that, given certain initial conditions, 
intraspecific competition and predation favor its development. These synchronous 
species should be favored by natural selection later to establish the 
life-cycles according the interspecific interactions or environment 
restrictions. 
Finally, it is established that the variable intensity of interspecific 
competition, predation and/or parasitism are interactions, which allow the 
existence of local stable positive periodical solutions for species such as 
insects and plants. However, as in any biological system, it is unlikely that 
one factor operates free from the influence of others.

\appendix 

\section{Proofs}

For the proofs,  we need some notations and definitions. 
We define the following quantities regarding the number of emergences in the interval $[1,T]$:
\begin{align*}
 (\mathcal{C}_1)& = \textrm{the number of times species } \mathcal{C}_1 \textrm{ emerges}, \\
 (\mathcal{C}_2)& = \textrm{the number of times species } \mathcal{C}_2 \textrm{ emerges},\\
 (\mathcal{C}_1 \wedge \mathcal{C}_2)& = \textrm{the number of times } \mathcal{C}_1 \textrm{ and } \mathcal{C}_2 \textrm{ emerge}\\ & \qquad \textrm{in the same year}, \\
 (\mathcal{C}_1 \wedge \overline{\mathcal{C}_2})& = \textrm{the number of times } \mathcal{C}_1 \textrm{ emerges and } \\ &\qquad \mathcal{C}_2 \textrm{ does not emerge},\\
 (\overline{\mathcal{C}_1} \wedge \mathcal{C}_2)& = \textrm{the number of times } \mathcal{C}_1 \textrm{ does not emerge }\\  &\qquad \textrm{ and }\mathcal{C}_2 \textrm{ emerges}.
\end{align*}
From $T=c_1\cdot c_2$  some elementary equalities follow:
\begin{align*}
(\mathcal{C}_1)=c_2,\nonumber \\
(\mathcal{C}_2)=c_1,\nonumber \\
(\mathcal{C}_1 \wedge \mathcal{C}_2)+(\mathcal{C}_1 \wedge \overline{\mathcal{C}_2})=c_2,\\
(\mathcal{C}_1 \wedge \mathcal{C}_2)+(\overline{\mathcal{C}_1} \wedge \mathcal{C}_2)=c_1.
\end{align*}

\subsection{Proof of Lemma 1}\label{ProofL1}
For the first six tuples we have
$$
F_1(c_1,c_2)=\frac{1}{c_2}(-(\mathcal{C}_1 \wedge \mathcal{C}_2))\leq \frac{-1}{c_2}<0.
$$
For the the last tuple we have
$$
F_2(c_1,c_2)=\frac{1}{c_1}(-(\mathcal{C}_1 \wedge \mathcal{C}_2))\leq \frac{-1}{c_1}<0.
$$

\subsection{Proof of Lemma 2}\label{ProofL2}

We will only give the proof for fixed points in the class $\mathcal{V}_4$, the other proofs are similar.
For the tuple $v_8=(1,1,-1,1)$ we have
$$
 F_1(c_1,c_2)=1, \quad
 F_2(c_1,c_2)=\frac{1}{c_1}((\overline{\mathcal{C}_1} \wedge \mathcal{C}_2)-(\mathcal{C}_1 \wedge \mathcal{C}_2)).
$$
For the tuple $v_9=(-1,1,1,-1)$ we have
\begin{align*}
 F_1(c_1,c_2)&=\frac{1}{c_2}(-(\mathcal{C}_1 \wedge \overline{\mathcal{C}_2})+ (\mathcal{C}_1 \wedge \mathcal{C}_2)), \\
 F_2(c_1,c_2)&=\frac{1}{c_1}((\overline{\mathcal{C}_1} \wedge \mathcal{C}_2)-(\mathcal{C}_1 \wedge \mathcal{C}_2)).
\end{align*}
For the tuple $v_{12}=(0,1,1,-1)$ we have
\begin{align*}
 F_1(c_1,c_2)&=\frac{1}{c_2}(\mathcal{C}_1 \wedge \mathcal{C}_2), \\
 F_2(c_1,c_2)&=\frac{1}{c_1}((\overline{\mathcal{C}_1} \wedge \mathcal{C}_2)-(\mathcal{C}_1 \wedge \mathcal{C}_2)).
\end{align*}
Therefore, 
\begin{align*}
 (c_1,c_2)\in \mathcal{A}_8 & \Leftrightarrow F_2(c_1,c_2')\geq F_2(c_1,c_2),\\
  (c_1,c_2)\in \mathcal{A}_9 & \Leftrightarrow F_1(c_1',c_2)\geq F_1(c_1,c_2) \wedge F_2(c_1,c_2')\geq F_2(c_1,c_2),\\
   (c_1,c_2)\in \mathcal{A}_{12} & \Leftrightarrow F_1(c_1',c_2)\geq F_1(c_1,c_2) \wedge F_2(c_1,c_2')\geq F_2(c_1,c_2).
\end{align*}
From that it is direct $\mathcal{A}_9=\mathcal{A}_{12}\subsetneq \mathcal{A}_8$. 
The strictness of the inclusion follows since, for instance, $(c_1,c_2)=(3,14)\in\mathcal{A}_8$, but 
$(c_1,c_2)=(3,14)\notin\mathcal{A}_9$ because 
$F_1(3+1,14) < F_1(3,14)$.

\subsection{Proof of Proposition 1}\label{ProofP1}

For $v_7=(1,1,-1,0)$ we have
\begin{align*}
 F_1(c_1,c_2)&=\frac{1}{c_2} ( (\mathcal{C}_1 \wedge \overline{\mathcal{C}_2}) 
 -(\mathcal{C}_1 \wedge \mathcal{C}_2) )=\frac{1}{c_2}(c_2-2(\mathcal{C}_1 \wedge \mathcal{C}_2)),\\
 F_2(c_1,c_2)&=\frac{1}{c_1} (\overline{\mathcal{C}_1} \wedge \mathcal{C}_2)= 
 \frac{1}{c_1}(c_1-2(\mathcal{C}_1 \wedge \mathcal{C}_2)).
\end{align*}
Clearly, if $\gcd (c_1,c_2)=1$, then
$$
 F_1(c_1,c_2)=\frac{c_2 - 2}{c_2},\quad F_2(c_1,c_2)=\frac{c_1 - 1}{c_1},
$$
and nobody may improve.

Consider the family
$$
(c_1,c_2)=(20,16^n\cdot 4\cdot 29\cdot 31)=(30,16^n\cdot 3596).
$$
Clearly, $\gcd(c_1,c_2)=2$, $\gcd(c_1-1,c_2)=29$, and $\gcd(c_1+1,c_2)=31$ so $F_1(c_1\pm 1,c_2)<F_1(c_1,c
_2)$, so the 
species $\mathcal{C}_1$ cannot improve its fitness.
Let us prove that the species $\mathcal{C}_2$ cannot change its period, as well. 
For that we will prove that $\forall n\geq 0$,  $5\mid 16^n\cdot 3596-1$ and $3\mid  16^n\cdot 3596+1$:
for $n=0$ clearly $3596-1=3593$ is divisible by $5$, and $3596+1=3597=3\cdot 1199$. 
The induction step for $c_2-1$ is as follows:
$$
16^{n+1}\cdot 3596-1=16^n \cdot 3596\times 16 -1 =16^n \cdot 3596\times 15 + (16^n \cdot 3596 -1).
$$
The first term on the right hand side is divisible by $5$, and the second term is divisible by $5$ by the 
induction hypothesis. 
Similarly, for $c_2+1$, we have
$$
16^{n+1}\cdot 3596+1=16^n \cdot 3596\times 16 +1 =16^n \cdot 3596\times 15 + (16^n \cdot 3596 +1).
$$
The first term on the right hand side is divisible by $3$, and the second term is divisible by $3$ by the 
induction hypothesis. 
Therefore, $\gcd(c_1,c_2+1)=5>\gcd(c_1,c_2)$ and $\gcd(c_1,c_2-1)=3>\gcd(c_1,c_2)$, so 
$F_2(c_1,c_2\pm 1)<F_2(c_1,c_2)$, hence all members of the family are local fixed points. 

Consider now $v_{10}=(-1,1,1,0)$. Consider, for example,
$$(c_1,c_2)=(30,30k+4),\quad k\geq 0.$$ Then $\gcd (c_1,c_2)=2$ so
\begin{align*}
 F_1(c_1,c_2)&=\frac{1}{c_2} ( -(\mathcal{C}_1 \wedge \overline{\mathcal{C}_2}) 
 +(\mathcal{C}_1 \wedge \mathcal{C}_2) )=\frac{-c_2 + 4}{c_2},\\
 F_2(c_1,c_2)&=\frac{1}{c_1} (\overline{\mathcal{C}_1} \wedge \mathcal{C}_2)= \frac{c_1 - 2}{c_1}.
\end{align*}
Since $c_1+1=31$ and $c_1-1=29$, $\gcd (c_1\pm 1,c_2)=1$, so the fitness of $\mathcal{C}_1$ cannot be improved. 
On the other hand, $c_2+1=30k+5$, so 
$\gcd(c_1,c_2+1)\geq 5$ and $F_2(c_1,c_2+1)<F_2(c_1,c_2)$. Similarly, $c_2-1=30k+3$, so 
$\gcd(c_1,c_2+1)\geq 3$ and $F_2(c_1,c_2-1)<F_2(c_1,c_2)$. Therefore, fitness of $\mathcal{C}_2$ cannot be improved, as well, so 
$(c_1,c_2)=(30,30k+4)$ is an fixed point for all $k\geq 0$. Furthermore, notice that if $c_2>4$, it is always $F_1(c_1,c_2)<0$, so the species 
$\mathcal{C}_1$ disappears. Also, if $(c_1,c_2)$ is such that $c_1\notin \{c_2-1,c_2+1\}$ and $c_2$ is a prime number, then $(c_1,c_2)$ 
is an fixed point and $\mathcal{C}_1$ also disappears.

Consider now $v_{11}=(1,0,-1,0)$. We have
\begin{align*}
 F_1(c_1,c_2)&=\frac{1}{c_2}( (\mathcal{C}_1 \wedge \overline{\mathcal{C}_2})- (\mathcal{C}_1 \wedge \mathcal{C}_2) ),\\
 F_2(c_1,c_2)&=0.
\end{align*}
If $\gcd (c_1,c_2)=1$, then $F_1=\displaystyle \frac{c_2-2}{c_2}$ is the maximum, and they are global fixed points. 
Further, couples $(c_1,c_2)=(30,30k+4)$, $k\geq 0$, are also local fixed points: because $\gcd (c_1,c_2)=2$, we have 
$$
\quad F_1(c_1,c_2)=\frac{1}{c_2}(2-(c_2-2))=\frac{-c_2+4}{c_2},\quad F_2(c_1,c_2)=0.
$$

For $v_{14}=(1,1,1,0)$ we have:
$$
  F_1(c_1,c_2)=1,\quad  F_2(c_1,c_2)=\frac{1}{c_1}(\overline{\mathcal{C}_1} \wedge \mathcal{C}_2).
$$
Clearly, all couples  $(c_1,c_2)$ such that $\gcd(c_1,c_2)=1$ are global fixed points, and, like in the previous case for $v_{10}$, 
the members of the family $(c_1,c_2)=(30,30k+4)$, $k\geq 0$, are local fixed points. 

For $v_{17}=(1,-1,-1,0)$, we have
\begin{align*}
 F_1(c_1,c_2)&=\frac{1}{c_2} ( (\mathcal{C}_1 \wedge \overline{\mathcal{C}_2}) 
 -(\mathcal{C}_1 \wedge \mathcal{C}_2) ),\\
 F_2(c_1,c_2)&=\frac{1}{c_1}(- (\overline{\mathcal{C}_1} \wedge \mathcal{C}_2)).
\end{align*}
Clearly, global fixed points do not exist: if $F_1(c_1,c_2)$ is at its maximum, that is, $(\mathcal{C}_1 \wedge \overline{\mathcal{C}_2})$ 
(for instance, if $\gcd(c_1,c_2)=1$), then $\mathcal{C}_2$ may reply by taking $c_2'=c_1$. 
Also, if $F_2(c_1,c_2)$ is at its maximum, that is $(\overline{\mathcal{C}_1} \wedge \mathcal{C}_2)=0$
(so $(\mathcal{C}_1 \wedge \mathcal{C}_2)=c_1$), then $\mathcal{C}_1$ may reply by taking $c_1'=c_2\pm 1$:
in that case $c_1',c_2)=1$ so $F_1(c_1',c_2)$ is better.

An infinite family of local fixed points is given by couples $(c_1,c_2)$ such that $c_2<c_1-1$ and $c_1=p$ is prime: then,
$$
F_1(p,c_2)=\frac{1}{c_2}((c_2-1)-1)=\frac{c_2-2}{c_2},
$$ which is at its maximum, and 
$F_2(p,c_2\pm 1)=F_2(p,c_2)$ because $p$ is prime so it is not divisible by any $c_2<p$.

The proofs for $v_{15}$, $v_{16}$ and $v_{18}$ are analogous.

\subsection{Proof of Proposition 2}\label{ProofP2}

For $v_1=(-1,0,1,1)$ we have
\begin{align*}
 F_1(c_1,c_2)&=\frac{1}{c_2} ( -(\mathcal{C}_1 \wedge \overline{\mathcal{C}_2}) + (\mathcal{C}_1 \wedge \mathcal{C}_2) ),\\
 F_2(c_1,c_2)&=\frac{1}{c_1} (\mathcal{C}_1 \wedge \mathcal{C}_2).
\end{align*}
If $c_1=c_2$, then $(\mathcal{C}_1 \wedge \overline{\mathcal{C}_2})=0$, so $F_1(c_1,c_2)=F_2(c_1,c_2)=1$ which is a maximum, 
so they are global fixed points. 

Consider now the family $(c_1,c_2)$ such that $c_1$ and $c_2$ are prime numbers and 
$c_1\notin \{c_2-1,c_2+1\}$ and $c_2\notin\{c_1-1,c_1+1\}$ (i.e., they are far enough). Then, 
$$
F_1(c_1,c_2)=\frac{1}{c_2}(-(c_2-1)+1)=\frac{-c_2+2}{c_2},\quad F_2(c_1,c_2)=\frac{1}{c_1}.
$$
Since $c_1$ and $c_2$ are primes, $c_1\pm 1$ and $c_2\pm 1$ are not divisors of $c_2$ or $c_1$, respectively, 
so fitness functions cannot improve. 

But, we have other families, too, which are not necessary primes, for instance: $(c_1,c_2)=(25,c_2)$, where
$$
 c_2\in \{ 7+10r \mid (r\geq 0) \wedge (3 \nmid 7+10r)\}= \{ 7+10r \mid r\geq 0\}\,\backslash\, \{ 27+30s \mid s\geq 0\}.
$$
The second set is is the set of numbers divisible by 3 and is contained in the first set since $27+30s=7+10(3s+2)$.
The difference set is infinite and, by Dirichlet's Theorem, since $\gcd(7,10)=1$, it contains infinitely many prime numbers.

For $v_{2}=(-1,1,1,1)$, we have
\begin{align*}
 F_1(c_1,c_2)&=\frac{1}{c_2} ( -(\mathcal{C}_1 \wedge \overline{\mathcal{C}_2}) 
 +(\mathcal{C}_1 \wedge \mathcal{C}_2) ),\\
 F_2(c_1,c_2)&=\frac{1}{c_1}c_1=1.
\end{align*}
If $c_2\mid c_1$, then $(\mathcal{C}_1 \wedge \overline{\mathcal{C}_2})=0$ so $F_1(c_1,c_2)=1$ which is the maximum so
it is a global fixed point. 

For $(c_1,c_2)=(2^p,2^q)$, $p<q$, it holds $\gcd(c_1\pm 1,c_2)=\gcd(2^p\pm 1,2^q)=1$, so
$$
F_1(c_1,c_2)=\frac{1}{2^q}(-(2^q-2^p)+2^p)=\frac{1}{2^q}(2^{p+1}-2^q),
$$
so it does not improve.

\subsection{Proof of Proposition 3}\label{ProofP3}

For $v_6=(1,1,-1,-1)$, we have
\begin{align*}
 F_1(c_1,c_2)&=\frac{1}{c_2}((\mathcal{C}_1 \wedge \overline{\mathcal{C}_2})- (\mathcal{C}_1 \wedge \mathcal{C}_2)), \\
 F_2(c_1,c_2)&=\frac{1}{c_1}((\overline{\mathcal{C}_1} \wedge \mathcal{C}_2)-(\mathcal{C}_1 \wedge \mathcal{C}_2)).
\end{align*}
If $\gcd(c_1,c_2)=1$, then
$$
F_1(c_1,c_2)=\frac{c_2-2}{c_2},\quad F_2(c_1,c_2)=\frac{c_1-2}{c_1},
$$
which is a global fixed point. 

The members of the family $(c_1,c_2)=(30,16^n\cdot 3596)$,
are also local fixed points - the proof is as for $v_7$ in Appendix \ref{ProofP1}.
%

\subsection{Proof of Proposition 4}\label{ProofP4}

For the tuple $v_8=(1,1,1,-1)$ we have
\begin{align*}
 F_1(c_1,c_2)&=1, \\ 
 F_2(c_1,c_2)&=\frac{1}{c_1}((\overline{\mathcal{C}_1} \wedge \mathcal{C}_2)-(\mathcal{C}_1 \wedge \mathcal{C}_2)). 
\end{align*}
Then $(c_1,c_2)$ such that $\gcd(c_1,c_2)=1$ is a global fixed point, since in this case
$(\mathcal{C}_1 \wedge \overline{\mathcal{C}_2})=(\overline{\mathcal{C}_1} \wedge \mathcal{C}_2)=1$, so
$$
F_2(c_1,c_2)=\frac{1}{c_1}((c_1-1)-1)=\frac{c_1-2}{c_1}
$$
is the maximum.

There exist infinitely many local fixed points, for instance 
$$(c_1,c_2)=(2\cdot 3^b\cdot5^c,4), \quad b,c\geq 1.$$
In this case $\gcd(c_1,c_2)=2$, so 
$$
F_2(c_1,c_2)=\frac{1}{c_1}((c_1-2)-2)=\frac{c_1-4}{c_1}.
$$
But, $c_2-1=3$, $c_2+1=5$,  and in both cases $\gcd(c_1,c_2\pm 1)\geq 3$, so 
$F_2(c_1,c_2\pm 1) <F_2(c_1,c_2)$.

Consider now $v_9=(-1,1,1,-1)$. We have
\begin{align*}
 F_1(c_1,c_2)&=\frac{1}{c_2}(-(\mathcal{C}_1 \wedge \overline{\mathcal{C}_2})+ (\mathcal{C}_1 \wedge \mathcal{C}_2)), \\
 F_2(c_1,c_2)&=\frac{1}{c_1}((\overline{\mathcal{C}_1} \wedge \mathcal{C}_2)-(\mathcal{C}_1 \wedge \mathcal{C}_2)).
\end{align*}
If $\gcd(c_1,c_2)=1$, then $(\mathcal{C}_1 \wedge \mathcal{C}_2)=1$ and 
$$
F_1(c_1,c_2)=\frac{-c_2+2}{c_2},\quad F_2(c_1,c_2)=\frac{c_1-2}{c_1}.
$$
If $c_2$ is prime and $c_1\pm 1\notin \{ c_2\pm 1\}$, then $(c_1,c_2)$ is a local fixed point. If, additionally, $c_2\geq 3$, 
the species $\mathcal{C}_1$ disappears.

But, there exist infinitely many local fixed points. Let us see one case: consider
$(c_1,c_2)=(2\cdot 3\cdot 5,2s)=(30,2s)$ such that $3 \mid (c_2-1)$ and $5\mid (c_2+1)$. Clearly, we have $\gcd(30,2s)=2$, so
\begin{align*}
F_1(c_1,c_2)&=\frac{1}{c_2}(-(c_2-2)+2)=\frac{-c_2+4}{c_2},\\
F_2(c_1,c_2)&=\frac{1}{c_1}((c_1-2)-2)=\frac{c_1-4}{c_1}.
\end{align*}
But, since $c_1-1=29$ and $c_1+1=31$ are both prime numbers, 
$$
F_1(c_1\pm 1,c_2)=\frac{1}{c_2}(-(c_2-1)+1)=\frac{-c_2+2}{c_2}< \frac{-c_2+4}{c_2}=F_1(c_1,c_2),
$$
so the fitness of the species $\mathcal{C}_1$ cannot improve. 

Let us see $c_2$: $2s\pm 1$ is not even and, 
by hypothesis, it is divisible by $3$ or $5$. Therefore, $\gcd(c_1,c_2\pm 1)\geq 3$ so
$F_2(c_1,c_2\pm 1)<F_2(c_1,c_2)$, and the fitness of the species $\mathcal{C}_2$ cannot 
improve either.

It remains to prove that there are infinitely many prime numbers $s$ such that 
$3 \mid (2s-1)$ and $5\mid (2s+1)$. Suppose 
$$
c_2-1=2s-1=3m,\quad c_2+1=2s+1=5n,\quad m,n\in \mathbb{N}.
$$
Then $3m+2=5n$, so 
$$
m=\frac{5n-2}{3}=\frac{(3+2)n-2}{3}=n+2\left(\frac{n-1}{3}\right), \quad n\in\mathbb{N}.
$$
There exist integer solutions for $m$ and $n$: for $n=4+3k$, $k\in\mathbb{N}$, $k\geq 0$ it holds
$$
m=\frac{5(4+3k)-2}{3}=\frac{18+15k}{3}=5k+6.
$$
Then 
$$
2s=3m+1=3(5k+6)+1=15k+19,\quad k\geq 0.
$$
Since $2s$ is even, $k$ must be odd. By setting $k=2l+1$, $l\geq 0$, we have
$$
c_2=2s\in \{34+30l \mid l\geq 0\}\equiv B.
$$
Set $B$ is an arithmetic progression such that $\gcd(17,15)=1$, hence by Dirichlet's Theorem it contains infinitely many primes. 
We conclude that there exist infinitely many couples $(30,2s)$, $s$ prime, which are local fixed points for the rule $v_9$.
Since $c_2>4$, we have $F_1(c_1,c_2)<0$, so the species $\mathcal{C}_1$ disappears. 
%
%
%

\bibliographystyle{spbasic}
\bibliography{Paper}

\end{document}